\documentstyle[12pt]{article}
\def\beq{\begin{equation}}
\def\eeq{\end{equation}}
\def\bea{\begin{eqnarray}}
\def\eea{\end{eqnarray}}
\def\bq{\begin{quote}}    
\def\eq{\end{quote}}

\def\bq{\begin{quote}}
\def\eq{\end{quote}}

\begin{document}

\baselineskip 24pt
\newcommand{\sheptitle}
{Atmospheric and Solar Neutrinos with a Heavy Singlet}

\newcommand{\shepauthor}
{S. F. King 
\footnote{On leave of absence from 
Department of Physics and Astronomy,
University of Southampton, Southampton, SO17 1BJ, U.K.}}

\newcommand{\shepaddress}
{Theory Division, CERN, CH-1211 Geneva 23, Switzerland.}

\newcommand{\shepabstract}
{We follow a minimalistic approach to neutrino masses, by introducing
a single heavy singlet $N$ into the standard model (or supersymmetric
standard model) with a heavy Majorana mass $M$, which couples as a
single right-handed neutrino in a Dirac fashion to leptons, 
and induces a single light see-saw mass $m_{\nu}\sim 5\times 10^{-2}\ eV$, 
leaving two neutrinos massless.
This trivial extension to the standard model
may account for the atomospheric
neutrino data via $\nu_{\mu}\rightarrow \nu_{\tau}$ oscillations
with near maximal mixing angle $\theta_{23}\sim \pi/4$ and 
$\Delta m_{\mu \tau}^2 \sim 2.5\times 10^{-3}\ eV^2$.
In order to account for the solar neutrino data
the model is extended to SUSY GUT/ string-inspired type models
which can naturally yield an additional light tau neutrino mass 
$m_{\nu_{\tau}}\sim few \times 10^{-3}\ eV$ leading to 
$\nu_{e L}\rightarrow (\cos \theta_{23}{\nu}_{\mu L}
-\sin \theta_{23}{\nu}_{\tau L})$  
oscillations with $\Delta m_{e1}^2\sim 10^{-5}\ eV^2$ and a mixing angle
$\sin^2 2 \theta_1 \approx 10^{-2}$
in the correct range for the small angle
MSW solution to the solar neutrino problem.
The model predicts
$\nu_{e L}\rightarrow (\sin \theta_{23}{\nu}_{\mu L}
+\cos \theta_{23}{\nu}_{\tau L})$  
oscillations with a similar angle but a larger splitting
$\Delta m_{e2}^2 \sim 2.5\times 10^{-3}\ eV^2 $.}

\begin{titlepage}
\begin{flushright}
CERN-TH/98-208\\
hep-ph/9806440\\
\end{flushright}
\begin{center}
{\large{\bf \sheptitle}}
\bigskip \\ \shepauthor \\ \mbox{} \\ {\it \shepaddress} \\ \vspace{.5in}
{\bf Abstract} \bigskip \end{center} \setcounter{page}{0}
\shepabstract
\begin{flushleft}
CERN-TH/98-208\\
\today
\end{flushleft}
\end{titlepage}

Atmospheric neutrino data from Super-Kamiokande \cite{SK} 
and SOUDAN \cite{SOUDAN}, when combined with the recent CHOOZ data
\cite{CHOOZ}, are consistent with
$\nu_{\mu}\rightarrow \nu_{\tau}$ oscillations
with near maximal mixing and 
$\Delta m_{\mu \tau}^2 \sim 2.5\times 10^{-3}\ eV^2$.
The data are equally consistent if $\nu_{\tau}$ is replaced by
a light sterile neutrino, and indeed many authors have considered
adding an extra light singlet neutrino state, in addition to the
usual three heavy singlet neutrinos which accompany the three
fermion families in quark-lepton unified models \cite{sterile}.
Without a light sterile neutrino, in the framework of some
unified theory, one is faced with the problem of specifying the
heavy Majorana matrix which is highly model-dependent and
requires additional assumptions. In many cases even with such additional
assumptions it is very difficult to account for atmospheric oscillations
within the framework of a complete description of quark and lepton masses
\cite{Dreiner}.\footnote{It is however rather straightforward to envisage
scenarios in which the large 23 mixing occurs via
the heavy Majorana matrix \cite{Allanach}. It remains to be seen
whether such approaches can be incorporated into complete models
however.}

In this paper we shall take a different tack. In order to overcome the
problem of the unknown and model-dependent heavy Majorana matrix, we
shall begin by assuming it does not exist. More precisely we shall
assume that the three ``right-handed neutrinos'' (or the equivalent
string-related physics possibly involving further singlet states
which can lead to low energy operators capable of giving
light effective Majorana masses)
have Majorana masses equal to the GUT scale $M_R\sim M_{GUT}$ leading to 
a third family neutrino mass $m_{\nu_{\tau}}\sim few \times 10^{-3} eV$,
with presumably smaller masses for the first and second families.
Since the atmospheric neutrino is heavier than these masses 
we shall ignore such masses to begin with. In order to account for the
atmospheric neutrino data we shall take a minimalistic approach
and introduce below the GUT scale only
a single heavy singlet $N$ into the standard model (or supersymmetric
standard model) with a heavy Majorana mass $M\bar{N}N^c$ where
$M<M_{GUT}$, which couples as a
single right-handed neutrino in a Dirac fashion to leptons, 
and induces a single light see-saw mass $m_{\nu}\sim 5\times 10^{-2}\ eV$, 
leaving two neutrinos massless.

To be exact the new singlet $N$ couples to the three lepton
doublets $L_{e,\mu ,\tau}$ and the Higgs doublet $H_2$
(the subscript follows the usual notation of the minimal supersymmetric
standard model, or MSSM, but the scheme also applies to the standard
model with a single Higgs doublet) as:
\beq
(\lambda_eL_e + \lambda_{\mu}L_{\mu}+\lambda_{\tau}L_{\tau})N^cH_2
\label{Ncoupling}
\eeq
where the subscripts $e,\mu ,\tau$ indicate that we are 
in the charged lepton mass eigenstate basis, e.g. $L_e=(\nu_{eL} , e^-_L)$
where $e^-$ is the electron mass eigenstate and $\nu_e$ is the associated
neutrino weak eigenstate, and $\lambda_e$ is a Yukawa coupling to the
singlet $N$ in this basis. When the Higgs develops its low
energy vacuum expectation value $v_2$ Eq.\ref{Ncoupling} will lead to
Dirac neutrino masses
\beq
v_2\bar{N}_R
(\lambda_e{\nu}_{eL} 
+ \lambda_{\mu}{\nu}_{\mu L}
+\lambda_{\tau}{\nu}_{\tau L})
\label{NDirac}
\eeq
corresponding to a single Dirac mass eigenstate
\beq
\nu_L=\frac{1}{\sqrt{|\lambda_e|^2+|\lambda_{\mu}|^2+|\lambda_{\tau}|^2}}
(\lambda_e{\nu}_{eL} 
+ \lambda_{\mu}{\nu}_{\mu L}
+\lambda_{\tau}{\nu}_{\tau L})
\label{nuL}
\eeq
with a Dirac mass
\beq
m_{\nu}^{Dirac}=\sqrt{|\lambda_e|^2+|\lambda_{\mu}|^2+|\lambda_{\tau}|^2}v_2
\eeq
Of course the singlet $N$ has a very large Majorana mass $M$, so
the physical mass is the light effective neutrino Majorana mass
given by the see-saw mechanism as:
\beq
m_{\nu}=\frac{{m_{\nu}^{Dirac}}^2}{M}=
(|\lambda_e|^2+|\lambda_{\mu}|^2+|\lambda_{\tau}|^2)\frac{v_2^2}{M}
\label{mnu}
\eeq
Clearly the light effective Majorana neutrino eigenstate
$\nu_L$ is the same as the Dirac neutrino eigenstate in Eq.\ref{nuL}.

In order to explain the atmospheric neutrino data 
we shall assume that the three Yukawa couplings,
which are completely free parameters in this model, satisfy the relation:
\beq
\lambda_e \ll \lambda_{\mu} \approx \lambda_{\tau}
\eeq
In this case the mass eigenstate neutrino can be expanded in terms
of the three weak eigenstate neutrinos in an approximate form
involving a small angle $\theta_{1}$ and a large angle $\theta_{23}$ as:
\beq
\nu_L=\theta_{1}{\nu}_{eL} 
+ s_{23}{\nu}_{\mu L}
+ c_{23}{\nu}_{\tau L}
\label{nuLmixing}
\eeq
which by comparison with Eq.\ref{nuL} implies
\beq
\theta_1 \approx \frac{\lambda_e}
{\sqrt{|\lambda_e|^2+|\lambda_{\mu}|^2+|\lambda_{\tau}|^2}}, \
s_{23} \approx \frac{\lambda_{\mu}}
{\sqrt{|\lambda_e|^2+|\lambda_{\mu}|^2+|\lambda_{\tau}|^2}}, \
c_{23} \approx \frac{\lambda_{\tau}}
{\sqrt{|\lambda_e|^2+|\lambda_{\mu}|^2+|\lambda_{\tau}|^2}}
\eeq
where $s_{23}=\sin \theta_{23}$, $c_{23}=\cos \theta_{23}$.

The interpretation of the atmospheric neutrino mixing is now clear.
There is a single massive neutrino $\nu_L$ with a Majorana
mass $m_{\nu}\sim 5\times 10^{-2}\ eV$, which contains large and 
approximately equal components of the weak eigenstates 
$\nu_{\mu L}$ and $\nu_{\tau L}$ parametrised by the 
mixing angle $\theta_{23} \sim \pi/4$. There will also be a small
admixture of the weak eigenstate $\nu_{eL}$ in the mass eigenstate
$\nu_L$, parametrised by the angle $\theta_1$
according to Eq.\ref{nuLmixing}. 
The two neutrino eigenstates which are orthogonal to $\nu_L$
will be massless. If we set $\theta_1 =0$ then one of the massless
neutrinos will be $\nu_{eL}$ and the other will be the
orthogonal combination 
\beq
\nu_{0L}\approx c_{23}{\nu}_{\mu L}-s_{23}{\nu}_{\tau L}
\label{nu0}
\eeq
In this approximation 
the atmospheric neutrino data is then consistent with
$\nu_{\mu}\rightarrow \nu_{\tau}$ oscillations via two state mixing
involving the massless neutrino $\nu_{0L}$ 
in Eq.\ref{nu0} and the massive neutrino 
$\nu_{L}$ in Eq.\ref{nuLmixing} with mass $m_{\nu}\sim 5\times 10^{-2}\ eV$.

So far we have provided a simple explanation of the atomospheric
neutrino data, assuming that $\theta_1 \ll \theta_{23} \sim \pi/4$.
We shall show that by including a single extra Majorana mass
for the tau neutrino $m_{\nu_{\tau}}\sim few \times 10^{-3} eV$
we can also account for the solar neutrino data via the
small angle MSW effect \cite{MSW}. 
Since $m_{\nu_{\tau}}\ll m_{\nu}$ the presence of such 
a mass in the $\nu_{\tau L}$ component of $\nu_L$ will only
slightly perturb the mass of this state at the 10 \% level.
However turning to the two orthogonal neutrinos, still assuming
that $\theta_1=0$, the state $\nu_{0L}$ in Eq.\ref{nu0}
will clearly pick up a mass of order 
$s_{23}^2m_{\nu_{\tau}}\sim m_{\nu_{\tau}}/2\sim 3\times 10^{-3} eV$ (say) 
via its $\nu_{\tau L}$ component, while $\nu_{eL}$ will remain massless.
When $\theta_1$ is switched on the weak
eigenstate $\nu_{eL}$ will consist dominantly of the massless state,
plus a small admixture of both the lighter and heavier mass eigenstates,
with the lighter of the two relevant for the
MSW solution of the solar neutrino problem.

We shall now discuss in more detail the structure of the
mixing matrix in the presence of a finite mixing angle $\theta_1$
and a non-zero tau neutrino mass $m_{\nu_{\tau}}$.
Taking $\theta_1\neq 0$ and small, 
but keeping $m_{\nu_{\tau}}=0$ to start with,
the mass spectrum of neutrinos consists of
one massive state $\nu_L$ in Eq.\ref{nuLmixing} with mass $m_{\nu}$
in Eq.\ref{mnu}, with two orthogonal combinations
$\nu_{eL} '$ and $\nu_{0L} '$ corresponding to two massless neutrinos. 
The charged weak currents in the
standard model are:
\beq
W_\mu^- ( \bar{e}, \bar{\mu}, \bar{\tau} )_L \gamma^\mu 
\left( \begin{array}{l} 
\nu_e\\
\nu_\mu\\
\nu_\tau
\end{array}
\right)_L
+ h.c.
\eeq
where $\nu_{eL}, \nu_{\mu L}, \nu_{\tau L}$ 
are neutrino weak eigenstates which couple with unit 
strength to $e$, $\mu$, $\tau$, respectively.
The mixing matrix $U$ is defined by the unitary transformation which relates
the weak eigenstates to the mass eigenstates:
\beq
\left(
\begin{array}{l}
\nu_e  \\
\nu_\mu  \\
\nu_\tau 
\end{array}
\right)_L = U 
\left(
\begin{array}{l}
\nu_e '\\
\nu_0 '\\
\nu
\end{array}
\right)_L
\label{defns}
\eeq
Since there is a degeneracy
in the massless neutrino subspace, the mixing matrix $U$ is only determined
up to an overall rotation in the two-dimensional massless subspace.
Since the degeneracy will ultimately
be lifted by $m_{\nu_{\tau}}$ (assuming that
the first and second family GUT generated masses are negligible)
it is convenient to rotate to a massless basis
where $\nu_{eL} '$ does not contain any
component of $\nu_{\tau L}$, and so will remain massless when
$m_{\nu_{\tau}}$ is eventually switched on.
In such a basis the mixing matrix is uniquely specified,
and to first order in $\theta_1$ is:
\beq
U = \left( \begin{array}{ccc}
 1 &  \left( \frac{c_{23}}{s_{23}}\right) \theta_1   & \theta_1     \\
-\frac{\theta_1}{s_{23}}   & c_{23} &  s_{23}   \\
0   &  -s_{23}   & c_{23} 
\end{array}
\right)
\label{U}
\eeq
In this basis the mass eigenstates expressed in terms of the weak
eigenstates are summarised below:
\begin{eqnarray}
\nu_{eL} ' &=& \nu_{eL} -\frac{\theta_1}{s_{23}} \nu_{\mu L} \nonumber \\
\nu_{0 L} ' & = & \left( \frac{c_{23}}{s_{23}}\right) \theta_1 \nu_{eL} 
+ c_{23} \nu_{\mu L} - s_{23}\nu_{\tau L} \nonumber \\
\nu_L &=& \theta_{1}{\nu}_{eL} 
+ s_{23}{\nu}_{\mu L}
+ c_{23}{\nu}_{\tau L}
\nonumber \\
\end{eqnarray}
In the $\theta_1=0$ limit $\nu_{eL} ',\nu_{0 L} '$
return to $\nu_{eL},\nu_{0 L}$ defined previously.

Now we consider the effect of a non-zero $m_{\nu_{\tau}}$.
Because of our judicious choice of basis the state $\nu_{eL} ' $
will remain massless because it does not have any tau neutrino component.
The only effect will be to re-mix the states $\nu_{0 L} '$ and
$\nu_L$, inducing a mass matrix in this sector of:
\beq
\left( \begin{array}{ll}
s_{23}^2m_{\nu_{\tau}} & -s_{23}c_{23}m_{\nu_{\tau}}     \\
-s_{23}c_{23}m_{\nu_{\tau}} & c_{23}^2m_{\nu_{\tau}}+m_{\nu} 
\end{array}
\right)
\eeq
Now $m_{\nu_{\tau}}/2\sim 3\times 10^{-3} eV$ 
is an order of magnitude smaller than $m_{\nu}\sim 5\times 10^{-2}\ eV$
so we expect additional mixing between $\nu_{0 L} '$ and
$\nu_L$ parametrised by a small angle
$\theta_{\tau}\sim s_{23}c_{23}\frac{m_{\nu_{\tau}}}{m_{\nu}}$.
With $m_{\nu_{\tau}}$ switched on 
we continue to denote the mass eigenstates as $\nu_{0L} ',\nu_L$ 
as in Eq.\ref{defns} with the modified mass eigenvalues:
\bea
m_{\nu_{0 L} '} & \approx & 
s_{23}^2m_{\nu_{\tau}} \sim 3\times 10^{-3} \ eV \nonumber \\
m_{\nu_L } & \approx & 
m_{\nu}+c_{23}^2m_{\nu_{\tau}} \sim 5\times 10^{-2}\ eV
\eea
leaving $\nu_{eL} ' $ massless. 
We continue to denote the mixing matrix as $U$
which is now modified from its form in Eq.\ref{U} as follows:
\beq
U = \left( \begin{array}{ccc}
 1 &  \left( \frac{c_{23}}{s_{23}}\right) \theta_1   & \theta_1     \\
-\frac{\theta_1}{s_{23}}   
& c_{23}(1+s_{23}^2 \frac{m_{\nu_{\tau}}}{m_{\nu}})
&  s_{23}(1-c_{23}^2 \frac{m_{\nu_{\tau}}}{m_{\nu}})   \\
0   
&  -s_{23}(1-c_{23}^2 \frac{m_{\nu_{\tau}}}{m_{\nu}})   
& c_{23}(1+s_{23}^2 \frac{m_{\nu_{\tau}}}{m_{\nu}}) 
\end{array}
\right)
\label{U'}
\eeq
The exact masses and mixing angles are given in the Appendix.

The physics of atmospheric neutrinos may be summarised as follows.
The weak eigenstate $\nu_{ \mu L}$ may be expanded in terms
of the mass eigenstates as:
\beq
\nu_{\mu L}  =   -\frac{\theta_1}{s_{23}}   \nu_{e L} ' +    
c_{23}(1+s_{23}^2 \frac{m_{\nu_{\tau}}}{m_{\nu}}) \nu_{0 L} ' 
 +   s_{23}(1-c_{23}^2 \frac{m_{\nu_{\tau}}}{m_{\nu}}) \nu_{ L}
\label{numuL}
\eeq
There will be atmospheric neutrino oscillations
$\nu_{\mu}\rightarrow \nu_{\tau}$ where maximal mixing 
now corresponds to $\tan \theta_{23}\approx 1+\frac{m_{\nu_{\tau}}}{m_{\nu}}$,
with 
\beq
\Delta m_{\mu \tau}^2  =  (m_{\nu}+m_{\nu_{\tau}}(c_{23}^2-s_{23}^2))^2
 \approx  m_{\nu}^2 \sim 2.5\times 10^{-3}\ eV^2
\label{mutau}
\eeq
The weak eigenstate $\nu_{\mu L}$ will also mix
slightly with $\nu_{e L} $ but since $\theta_1 \ll \theta_{23}$
this mixing will not be noticed by the atmospheric neutrino experiments. 

The situation regarding $\nu_{eL}$ mixing is quite interesting.
The weak eigenstate $\nu_{eL}$ may be expanded in terms
of the mass eigenstates as:
\beq
\nu_{e L}  = \nu_{e L} ' 
+ \left( \frac{c_{23}}{s_{23}}\right) \theta_1 \nu_{0 L} '
+ \theta_1 \nu_{ L}
\label{nueL}
\eeq
An electron neutrino weak eigenstate $\nu_{e L}$ 
will therefore have a dominant massless component $\nu_{e L} ' $
which is split from the massive states by two different mass
splittings:
\bea
\Delta m_{e1}^2 & = &  (m_{\nu_{0 L} '} -m_{\nu_{eL} '})^2=m_{\nu_{0 L} '}^2
\sim 10^{-5}\ eV^2 \nonumber \\
\Delta m_{e2}^2 & = & (m_{\nu_L }  -m_{\nu_{eL} '})^2 = m_{\nu_L}^2
\sim 2.5\times 10^{-3}\ eV^2
\label{Delta12}
\eea
with the physics of electron neutrino oscillations governed by
Eqs.\ref{nueL} and \ref{Delta12}. Electron neutrino oscillations 
can be roughly characterised as follows:
\beq
\nu_{e L}\rightarrow \nu_{0 L} '\approx
c_{23}{\nu}_{\mu L}-s_{23}{\nu}_{\tau L}
\eeq
with 
$\Delta m_{e1}^2 $ and a mixing
amplitude $\left( \frac{c_{23}}{s_{23}}\right) \theta_1$.
In addition we expect oscillations to the orthogonal combination:
\beq
\nu_{e L}\rightarrow \nu_L \approx s_{23}{\nu}_{\mu L}+c_{23}{\nu}_{\tau L}
\eeq
with $\Delta m_{e2}^2$ and a mixing
amplitude $\theta_1 $. Assuming that $s_{23} \approx c_{23}$, then
the solar neutrino problem may be solved by the small angle MSW solution with
$\sin^2 2\theta_{1}\sim 10^{-2}$, in which case the oscillations with
$\Delta m_{e1}^2 \sim 10^{-5}\ eV^2 $ will be resonantly enhanced.
The orthogonal oscillations with $\sin^2 2\theta_{1}\sim 10^{-2}$ and 
$\Delta m_{e2}^2 \sim 2.5\times 10^{-3}\ eV^2 $ are not resonantly 
enhanced and play only a negligible role in solar physics,
however eventually one might hope to observe such oscillations
in {\it vacuo}. The new generation of solar neutrino experiments
will be able to measure the mixing angle $\theta_{23}$ by 
counting the number of ${\nu}_{\mu L}$ and ${\nu}_{\tau L}$ events
which will be equal in the case of maximal atmospheric mixing.
 
We now give a brief comparison to recent approach 
which has some superficial similarities to our scheme \cite{Drees}. 
The addition of a single
gauge singlet $N$ with Yukawa couplings in Eq.\ref{NDirac}
gives rise to a light effective Majorana matrix in the
$\nu_{e L} , \nu_{\mu L}, \nu_{\tau L}$ basis proportional to 
\beq
\left( \begin{array}{lll}
 \lambda_e^2 &  \lambda_e \lambda_{\mu} & \lambda_e \lambda_{\tau}     \\
\lambda_e \lambda_{\mu} & \lambda_{\mu}^2 & \lambda_{\mu} \lambda_{\tau} \\
\lambda_e \lambda_{\tau} & \lambda_{\mu} \lambda_{\tau} & \lambda_{\tau}^2
\end{array}
\right)\frac{v_2^2}{M}
\label{Maj}
\eeq
assuming for simplicity that the Yukawa couplings are real.
Such a matrix is similar to that proposed recently \cite{Drees},
and clearly gives rise to one massive neutrino and 
two massless neutrinos. We have simply assumed that
$\lambda_e \ll \lambda_{\mu} \sim \lambda_{\tau}$ in order to account
for the atmospheric neutrino results, but clearly such a structure
could be enforced by suitable (approximate) discrete symmetries.
However whereas in our model the basic structure in Eq.\ref{Maj}
is a simple consequence of
having a single gauge singlet, in the other model it is due to
R-parity violating couplings with certain discrete symmetries \cite{Drees}. 
We have shown that a small perturbation in the 33 element
of the matrix in Eq.\ref{Maj} coresponding to a tau neutrino light Majorana
mass of $m_{\nu_{\tau}}$ coming from GUT or string scale physics
can give a small mass to one of the previously massless eigenstates,
leaving the other one massless, and have used this small splitting
to account for the solar neutrino data. There will of course be other
contributions to the mass matrix coming from GUT or string scale
masses involving the first and second families, but they are expected
to be much smaller than $m_{\nu_{\tau}}$ according to general arguments,
and we have ignored them here. By contrast the perturbations to the
other elements of the matrix in \cite{Drees} are comparable to the
33 perturbation, with large model dependence. This implies that our
simple predictions for electron neutrino oscillations 
discussed above will not
be shared by the other model \cite{Drees}.

As regards the origin of the singlet $N$
we remark that our model may be embedded into an $SO(10)$ type model
if the singlet $N$ originates from some new $16 '$ which is part of a vector
representation $16 ' + \bar{16} '$ with a heavy vector mass 
$M_V 16 ' \bar{16} '$ which may be comparable to the
Majorana mass $M 16 ' 16 ' \rightarrow M N N$. The standard 
couplings $16_i 16_j 10$ are then supplemented by new couplings 
$16_i 16 ' 10$ which include the Dirac neutrino couplings to the
singlet $N$ contained in the $16 '$. It will also include additional
Dirac couplings which mix charged quarks and leptons with the
new exotic quarks and leptons contained in the $16 '$. However the
new exotic quarks and leptons all have large vector masses $M_V$, so such
couplings give a negligible contribution to
charged quark and lepton masses and mixing angles.
The standard ``right-handed neutrinos'' contained in the $16_i$'s
are all assumed to have GUT scale masses in such a scenario,
leading to the small tau neutrino mass perturbation that we
need for the solar neutrino problem. There may of course be other 
origins for the singlet $N$ which have nothing to do with $SO(10)$.

Finally we remark that with our neutrino spectrum consisting of two
neutrinos of mass $5\times 10^{-2}\ eV$ and $3\times 10^{-3} eV$ and 
one massless (or strictly very light) neutrino, it is not possible 
to use this model to account for hot dark matter.
We also cannot account for the LSND data \cite{LSND}.
If either becomes mandatory then the scheme described here would
be excluded.

\begin{center}
{\bf Acknowledgements}
\end{center}
I would like to thank Steve Abel, Sacha Davidson and Graham Ross
for useful comments.

\vspace{0.5in}

\begin{center}
{\bf Appendix: Exact Masses and Mixing Angles}
\end{center}
There are 5 underlying parameters as defined in the text:
\beq
\lambda_e, \ \lambda_{\mu}, \ \lambda_{\tau}, \ M, m_{\nu_{\tau}}
\eeq
which are assumed to all take real values.
From these 5 parameters we obtain 4 physical observables,
consisting of 2 masses $m_{\nu_{\tau}}$ and
$m_{\nu}$ defined in Eq.\ref{mnu}, plus
2 angles defined as:
\beq
s_1\equiv \sin \theta_1 \equiv \frac{\lambda_e}
{\sqrt{|\lambda_e|^2+|\lambda_{\mu}|^2+|\lambda_{\tau}|^2}}
\eeq
\beq
t_{23}\equiv \tan \theta_{23} \equiv \frac{\lambda_{\mu}}{\lambda_{\tau}}
\eeq
The exact expressions for neutrino masses and mixing angles are then 
expressed in terms of $\theta_1 , \theta_{23}, m_{\nu} , m_{\nu_{\tau}}$.

The mixing matrix $U$ is defined by the unitary transformation which relates
the weak eigenstates to the mass eigenstates:
\beq
\left(
\begin{array}{l}
\nu_e  \\
\nu_\mu  \\
\nu_\tau 
\end{array}
\right)_L = U 
\left(
\begin{array}{l}
\nu_e '\\
\nu_0 '\\
\nu
\end{array}
\right)_L
\label{defnsp}
\eeq
The exact mass eigenvalues are:
\bea
m_{\nu_{eL} '} & = & 0 \nonumber \\
m_{\nu_{0 L} '} & = & 
\delta_{11}c_{\tau}^2-2\delta_{12}c_{\tau}s_{\tau}+(\delta_{22}+m_{\nu})
s_{\tau}^2 
\nonumber \\
m_{\nu_L } & = &
(\delta_{22}+m_{\nu})c_{\tau}^2 
+2\delta_{12}c_{\tau}s_{\tau} + \delta_{11}s_{\tau}^2
\eea
where
\bea
\delta_{11}& = & s_{23}^2(1+\frac{s_1^2}{t_{23}^2})^{1/2}m_{\nu_{\tau}}
\nonumber \\
\delta_{22}& = & c_{23}^2c_1^2m_{\nu_{\tau}}
\nonumber \\
\delta_{12}& = & s_{23}c_{23}c_1(1+\frac{s_1^2}{t_{23}^2})^{1/2}m_{\nu_{\tau}}
\label{delta}
\eea
and $s_{\tau}=\sin \theta_{\tau}$, $c_{\tau}=\cos \theta_{\tau}$
where $\theta_{\tau}$ is the additional mixing angle induced by
$m_{\nu_{\tau}}$ and is given by
\beq
\tan 2 \theta_{\tau}=\frac{2 \delta_{12}}{m_{\nu}+\delta_{22}-\delta_{11}}
\eeq
The exact mixing matrix is:
\beq
U = \left( \begin{array}{ccc}
\left(1+\frac{t_1^2}{s_{23}^2}\right) ^{-\frac{1}{2}}
&  \frac{c_{\tau}s_1}{t_{23}}
\left(1+\frac{t_1^2}{s_{23}^2}\right) ^{-\frac{1}{2}} +s_{\tau}s_1
& c_{\tau}s_1 -   \frac{s_{\tau}s_1}{t_{23}}
\left(1+\frac{t_1^2}{s_{23}^2}\right) ^{-\frac{1}{2}} \\

-\frac{t_1}{s_{23}}\left(1+\frac{t_1^2}{s_{23}^2}\right) ^{-\frac{1}{2}}
&  c_{\tau}c_{23}c_1^2
\left(1+\frac{s_1^2}{t_{23}^2}\right) ^{-\frac{1}{2}} +s_{\tau}s_{23}c_1
& c_{\tau}s_{23}c_1 -  s_{\tau}c_{23}c_1^2 
\left(1+\frac{s_1^2}{t_{23}^2}\right) ^{-\frac{1}{2}} \\
0   
&  -c_{\tau}s_{23}
\left(1+\frac{s_1^2}{t_{23}^2}\right) ^{\frac{1}{2}} +s_{\tau}c_{23}c_1
& c_{\tau}c_{23}c_1 +  s_{\tau}s_{23} 
\left(1+\frac{s_1^2}{t_{23}^2}\right) ^{\frac{1}{2}}
\end{array}
\right)
\label{U''}
\eeq

\end{document}